\documentstyle[aas2pp4,amstex,amsfonts,epsfig,rotating,float]{article}

\def\taut{\tau_{\rm T}}

\def\refindent{\par\noindent\hangindent=3pc\hangafter=1 }
\def\aa#1#2#3{\refindent#1, A\&A, #2, #3}
\def\aasup#1#2#3{\refindent#1, A\&AS, #2, #3}

\def\apj#1#2#3{\refindent#1, {\it ApJ}, {\bf#2}, #3.}
\def\apjlett#1#2#3{\refindent#1, {\it ApJ (Letters)}, {\bf #2}, #3.}
\def\apjsup#1#2#3{\refindent#1, ApJS, #2, #3}

\def\mnras#1#2#3{\refindent#1, {\it MNRAS}, {\bf#2}, #3.}
\def\nature#1#2#3{\refindent#1, {\it Nature}, {\bf #2}, #3.}

\def\pasp#1#2#3{\refindent#1, PASP, #2, #3}

\begin{document}
\centerline{Submitted to the Astrophysical Journal}
\bigskip
\title{Physical Constraints on, and a Model for, the Active\\
    Regions in Seyfert Galaxies}

\author{Sergei Nayakshin$^*$ and Fulvio Melia$^{*\dag}$\altaffilmark{1}}
\affil{$^*$Physics Department, University of Arizona, Tucson, AZ 85721}
\affil{$^{\dag}$Steward Observatory, University of Arizona, Tucson, AZ 85721}

\altaffiltext{1}{Presidential Young Investigator.}






\begin{abstract}
We discuss several physical constraints on the nature of
the Active Regions (AR) in Seyfert 1 Galaxies, and show
that a plausible model consistent with these constraints 
is one in which the ARs are magnetically confined and ``fed''. 
The unique X-ray index of these sources points to a
large compactness parameter ($l\gg 1$).  This, together with
the conditions required to account for the observed
optical depth being close to unity, suggests that
the magnetic energy density in the AR should be
comparable to the equipartition value in 
the accretion disk, and that it should be released 
in a flare-like event above the surface of the cold
accretion disk.  We consider the various issues pertaining
to magnetic flares and attempt to construct a coherent picture,
including a reason for the optical depth in the AR being $\sim 1$, and
an understanding of the characteristics of the X-ray reflection
component and the power density spectra associated with this
high-energy emission.
\end{abstract}


\keywords{acceleration of particles --- black hole physics --- magnetic
fields --- plasmas --- radiation mechanisms: non-thermal --- galaxies: Seyfert}



%

\section{Introduction}
X-ray emission is a major contributor to the observed spectrum of 
Seyfert Galaxies, and yet the physics of the emitting region 
is still not well understood. The most common and (thus far) successful 
approach to this problem, to which we shall refer as the `spectral approach' (SA),
makes very simple assumptions about the geometry and/or the
particle heating mechanisms, but uses a detailed microphysical approach to
account for the particle-photon interactions and to derive the spectrum.
This spectrum is then compared with the observations in order 
to place constraints on the parameters of the emitting regions. 
The early models assumed a non-thermal pair dominated plasma. 
(For a comprehensive review of non-thermal models see Svensson 1994.) 
However, with the more recent substantial progress made in the X-ray
observation of Seyfert Galaxies (e.g., Jourdain et al. 1992; 
Johnson et al. 1993), it is now evident that thermal models are 
strongly favored by the data.  Accordingly, much of the current
attention is focused on thermal models (Svensson 1996a).

Aside from the question concerning the nature of the particle distribution, 
there is also the issue regarding the emitter's geometry. Haardt \& 
Maraschi (1991, 1993) argued that if most of the energy is dissipated 
in a hot corona overlying a cold accretion disk, then the resulting 
spectrum naturally explains many of the observed features in these
sources. In particular, roughly half of the coronal X-ray emission
is directed towards the cold disk, where it gets absorbed and re-emitted
as UV radiation, which then re-enters the corona and contributes to
the cooling of the electrons. Thus, the lepton cooling rate becomes 
proportional to the heating rate.  In this case, the inverse Compton 
up-scattering of the UV radiation leads to an almost universal X-ray 
spectral index, consistent with the observations (e.g., according to 
Nandra \& Pounds 1994, $\alpha \simeq 1.95\pm 0.15$ for a sample of 
Seyfert Galaxies).  The hardening of the spectrum above about 10 keV 
(Nandra \& Pounds 1994) and a broad hump at $\sim$ 50 keV (e.g., 
Zdziarski et al 1995) are accounted for by reflection of the 
hard X-rays in the cold disk.

However, observationally the hard X-ray luminosity, $L_h$, can 
be a few times smaller than the luminosity, $L_s$, in the soft UV-component. 
This is inconsistent with the uniform two-phase disk coronal model, 
because the latter predicts about the same luminosity in both X-rays 
and UV (due to the fact that all the UV radiation arises as a consequence 
of reprocessing of the hard X-ray flux, which is about equal in the
upward and downward directions).  To overcome this apparent difficulty, 
Haardt, Maraschi \& Ghisellini (1994) introduced a patchy disk-coronal 
model, which assumes that the X-ray emitting region consists of separate 
\lq active regions\rq (AR) independent of each other. In this case, a portion
of the reprocessed as well as intrinsic radiation from the cold disk escapes
to the observer directly, rather than entering ARs, thus allowing for a
greater ratio of $L_s/L_h$. 

Recently, Stern et al. (1995) and Poutanen \& Svensson (1996a) carried out 
state of the art calculations of the radiative transport of the 
anisotropic polarized radiation, for a range of AR geometries. They
showed that this type of model indeed reproduces the observed 
X-ray spectral slope, the compactness, and the high-energy cutoff.
The model has very few parameters, namely, the compactness 
and the temperature of the intrinsic/reprocessed radiation from the cold disk. 
Therefore, it appears that the model is very robust in its predictions.

On the other hand, another somewhat less common approach to explaining
the X-rays from Galactic black hole candidates (GBHC) and Seyfert Galaxies, which 
we shall call the `magnetic flare' (MF) approach, is being developed by
analogy with the strong, energetic (X-ray emitting) flares observed on the 
Sun. A pioneering paper in this field was that of Galeev, Rosner \& 
Vaiana (1979), who showed that the physical conditions in an accretion 
disk surrounding a black hole are such that magnetic fields
are likely to grow to equipartition values. This magnetic field 
is then transported to the surface of the disk by buoyancy forces
where its energy is released in a flare-like event. 
The magnetic flare approach is, in a sense, complementary to the spectral
approach, in that it attempts to include all the relevant physics 
self-consistently (e.g., de Vries \& Kuijpers 1992; van Oss, 
van den Oord \& Kuperus 1993; Volwerk, van Oss \& Kuijpers 1993). 
Unfortunately, the physics involved is quite complex and still 
somewhat open to debate.  The resulting spectrum is a combination 
of time-averaged components from many different flares, and is subject to
many uncertainties---clearly the MF model must invoke many more 
parameters, or assumptions about the magnetic field reconnection,
than does the SA approach. Therefore, no detailed spectra 
from these events (in the case of Seyfert Galaxies) have yet been computed.

One may argue that to make substantial progress, these two approaches
need to find an overlap of self-consistency. In particular, the SA model
does not specify the mechanism by which the gravitational energy dissipated
within the cold disk is transported out to the optically thin corona.
It is {\it assumed} that some process can provide the needed electron 
heating, and often a reference is made to magnetic fields.  Moreover, 
the particle dynamics is ignored, imposing instead the artificial 
constraint that the particles are confined to a closed box. Thus, even though
the SA model can reproduce the observed spectrum quite well, the situation
is unsatisfactory from a broader theoretical perspective. 
 
Correspondingly, it appears that the most important results obtained
within the framework of the SA model have not been fully incorporated
into the magnetic flare scenarios. For example, it is well known that the 
universal X-ray spectral index in Seyfert Galaxies is
best explained by the inverse Comptonization of soft UV photons. This requires 
a relatively high value of the compactness parameter (see discussion below) 
in the emitting region.  As far as we can tell, no work has yet been done to
show (based on the physics of reconnection or some other mechanism for
the transfer of energy from the magnetic field to the particles)
that a specific MF model can indeed provide the 
needed high compactness during the active phase, though Haardt, Maraschi 
\& Ghisellini (1994, hereafter HMG94) did use the physics of magnetic
flares to account for the heating rates and the required confinement
of the ARs. They showed that the compactness can be high enough during 
the active phase if one assumes that the entire magnetic field energy 
is transfered to the particles during a few light-crossing time scales.
They did not, however, explicitly consider the question of
how the spectrum from these highly transient phenomena is formed.

More recently, Nayakshin \& Melia (1997b) considered the issue of pressure
balance within the plasma trapped inside the flare during the active phase.
They found that under certain conditions, a pressure equilibrium 
can be maintained in the source if its Thomson optical depth 
is $\tau_T = 1 - 2$. They also showed that the current data cannot
distinguish between a spectrum comprised of a single flare component
and one formed from many different flares with a range $\sim 0.5 -2$ 
in $\tau_T$. In other words, one can always find a $\tau_T$ for the 
spectrum assuming a single flare that represents the composite 
spectrum quite well out to about 100 keV, where the quality of the
data deteriorates. In addition, Nayakshin \& Melia (1997c) have
considered the implications of a time-dependent X-ray reflection and
reprocessing by the cold disk underneath the flare. They find that 
due to the short lived, but very intense X-ray flux from the AR, 
the upper layer of the disk is compressed to a density in excess
of that found in the disk's mid-plane.  Under these conditions, the 
X-ray reprocessing leads to a temperature of the emitted
UV radiation that is roughly independent of the X-ray luminosity and the 
overall bolometric luminosity of the source, as suggested by the 
EUV-soft X-ray observations (Walter \& Fink 1993; Fink et al. 1994; 
Zhou et al. 1997). Due to the increased gas density in the compressed 
layer, the ionization parameter is smaller than that arising in 
time-independent X-ray reflection (i.e., when the X-ray source is 
assumed to be stationary---a condition that is clearly violated in
magnetic flares).  This may explain those observations of Seyfert
Galaxies that suggest the presence of a nearly neutral reflector
(Zdziarski et al. 1996). 

These results strengthen the MF model and motivate us here to
attempt to assemble the various components of this picture. We first discuss 
the different physical constraints imposed on the ARs
by both the spectral observations and the physics of the corresponding
processes, without necessarily confining our discussion to the MF model.
We will then show that magnetic flares above the cold disk are probably
the best candidates for producing these ARs, and we discuss the
physics of the MF model in greater depth. We conclude by listing
some of the unresolved issues.

\section{Physical Constraints on the Active Regions}

Our first task here will be to assemble the various constraints
imposed on the ARs in Seyfert Galaxies from observations and theoretical
considerations. In so doing, we shall first summarize the better known
results, and then discuss the additional constraints that follow 
from various attempts to construct realistic ARs based on the
idea that these may be magnetic structures, characterized by
a well-defined confinement and energy supply.

\subsection{Compactness of the Active Regions}

The most important parameter of the AR is the compactness 
$l\equiv F_{\gamma}\sigma_T \Delta R/m_e c^3$,
where $F_{\gamma}$ is the radiation energy flux at the top of the AR
and $\Delta R$ is its typical size.
Note that this definition is for the local compactness, i.e., the one
that characterizes the local properties of the plasma, unlike the global compactness
$l_g\equiv L\sigma_T/R' m_e c^3$, where $L$ is the total luminosity
of the object and $R'$ is the typical size of the region that emits this luminosity.
It is the latter that should be compared to the observed compactness rather 
than the former. 

Consider the following example. Assume that the emitting region
is a full disk-like corona. In this case the local and global compactnesses
are related in this way:
\begin{equation}
l\simeq l_g\, {H_c\over R'}\;,
\end{equation}
where $H_c$ is the coronal scale height, which is unlikely
to be larger than the accretion disk scale height, $H_d$, and so the local
compactness $l$ can be much smaller than the global one. At the same time,
as suggested by the frequently observed large ratio $L_s/L_h$ (e.g., HMG94; 
Svensson 1996a), the emitting region can consist of a large number of small localized
areas.  Since the total X-ray luminosity from these ARs should be
the same as that in the model with a full corona, it is clear that
the local compactness of each region must be larger than that of the full
corona. In particular, depending on the ratio of the total active
area $\Delta S$ covered by the ARs to ${R'}^2$, the local compactness
can be either larger or smaller than the global one.
Therefore, even though the observed values of global compactness for 
Seyfert Galaxies lie in the range $1-100$ (Done \& Fabian 1989; but
see also Fabian 1994), one cannot argue that the local compactness $l$
should be larger than these values based on observations alone.

However, a large local compactness is strongly
preferred in current pair-dominated two-phase models (e.g., 
Svensson 1996a; Zdziarski et al. 1996). To produce the correct spectrum, 
$\taut$ should be relatively large ($\sim 1$). In the context
of the pair-dominated two-phase model, the only mechanism
for fixing the optical depth is by pair equilibrium,
and thus one needs $l \gg 1$ in order to create them,
in which case the optical depth of the ARs becomes a function of compactness.
However, as we show in \S 3.3 below, the optical depth of the X-ray 
emitting regions may be dominated by electrons rather than pairs. 
For the purposes of setting theoretical limits on the compactness 
parameter, this nevertheless implies the same result since
both cases require $l\gg 1$.

In addition, radiation mechanisms put their own limitations on the local
compactness. The fact that the X-ray spectral index for Seyfert
Galaxies lies in a rather narrow range (Nandra \& Pounds 1994) is
most naturally explained by the approximately constant Compton
$y$-parameter (defined in, e.g., Rybicki \& Lightman 1979). 
Fabian (1994) shows that in order for the Compton emissivity to
dominate over the bremsstrahlung one, the compactness of the plasma should be
larger than 
\begin{equation}
l\sim 0.04 \Theta^{-3/2}\;,
\end{equation}
where $\Theta$ is the electron temperature in the units of $m_e c^2/k_B$.
For the typical value $\Theta\sim 0.2$, this requires that $l \gtrsim 0.5$.

Note that the gas does not necessarily need to be Maxwellian, 
as long as the optical depth is sufficiently large (e.g., Ghisellini
et al. 1994; Nayakshin \& Melia 1997a), since then the Comptonized
spectrum looks very much the same for different electron distributions
having the same $y$-parameter. Moreover, in the presence of a strong
magnetic field, the synchrotron self-absorption is an efficient
mechanism for thermalizing the electrons, to the extent that it 
becomes a more important thermalization mechanism than Coulomb 
collisions (Svensson 1996b; Nayakshin \& Melia 1997a). Thus, constraints 
imposed on the compactness by the Coulomb thermalization process 
(Fabian 1994) can be violated.

To summarize this section, we note that all current
explanations for the X-ray emission from Seyfert Galaxies require a
large local compactness parameter $l\gg 1$.

\subsection{Geometry, Confinement and Life Time: Magnetic Flares 
Required}

As already discussed in the Introduction, observational evidence 
very strongly favors the geometry 
of localized X-ray sources above the accretion disk. We note that
this immediately requires the active regions to be transient with a
lifetime comparable to (or less than) the disk thermal time scale, 
$t_{\rm th}\sim \alpha^{-1}t_{\rm h}$, where
$t_{\rm h} = H_d/c_s$ in terms of the sound speed $c_s$. An
integral assumption of the two-phase model is that the internal disk
emission is negligible compared with the X-ray flux of the AR, at least
during the active phase (Poutanen \& Svensson 1995). Assuming that a 
fraction ($\sim 1$) of the total energy content in the surface
area of the disk immediately below the AR is transferred into the AR, 
the time scale for the release of this energy must then be much 
shorter than $t_{\rm th}$, during which time the disk's internal energy 
is radiated. Our calculations
show that if this condition is not satisfied, then the localized ARs actually
produce a {\it steeper} spectrum than that of a full corona, due to the enhanced 
internal emission from regions of the disk that surround the AR.  This
is an effect that is neglected in the two-phase corona-disk model.
Physically, the internal disk emission provides too much cooling in this case,
unless the X-ray emitting region somehow snatches heating power even from disk
regions that are not directly below it, which appear to be unrealistic.

The plasma in
the ARs should be confined during the active phase, otherwise the energy
will be lost to the expansion of the plasma rather than producing the X-rays.
Not confined, the source would expand at the sound speed (which turns out 
to be a fraction of $c$ for these conditions). The lifetime
of the AR would then be limited to a few light crossing times. It is not
clear that the spectrum from such an expanding and short lived source
can resemble anything studied thus far in the literature.
The familiar gravitational confinement, operating in the main part of the
accretion disk, does not work here due to several reasons. First of all,
the {\it locally} limited Eddington compactness $l$ is at most $\sim 50/(1 + 2z)$
for ARs with a roughly semi-spherical shape,
where $z$ is the positron number density $n_+$ divided by that of the protons 
$n_p$, while the relatively large Thomson optical depth $\tau_T\equiv
\sigma_T n_p (1 + 2 z)\sim 1$ obtained by Zdziarski et al. (1996)
requires a compactness of a few hundred (if no magnetic field is involved 
and the particles are confined to a rigid box). Second, 
there is no mechanism for counter balancing a side-ways expansion of the 
plasma. Therefore, since there seems to be no other reasonable 
possibility for confinement of the AR plasma, it may be argued that
a magnetic field is required to provide the bounding pressure.
Any confinement mechanism will fail to confine the
plasma for a time longer than about one dynamical time scale for the disk,
since adjacent points with slightly different radii are torn apart
on this time scale due to the disk's differential rotation. 

In addition, if the pairs are important for the model, then the lifetime
of the AR should be large enough to allow establishment of the pair 
equilibrium. To put it another way, there should be enough time
to create enough pairs if the plasma is initially optically thin and 
proton-dominated. We experimented with time-dependent codes in which
radiation transfer is treated in the frequency-dependent 
Eddington approximation, and found that this condition leads to
the requirement that the lifetime of the region should be roughly
an order of magnitude longer than the light crossing time for the AR.
In the thin disk approximation one can always satisfy both
requirements as long as the size $\Delta R$ of the AR is of the order of
the disk scale height $H_{\rm d}$, since $t_{\rm h}
\equiv  H_{\rm d}/c_s \gg H_{\rm d}/c$, where $c_s$ is the local sound speed. 
We should also note that there can be other than pair creation 
mechanisms for the plasma to adjust its optical depth (see \S 3.3), so
this constraint is only important when the optical depth is dominated
by pairs.

To be consistent with the observations and the physics of the two-phase
accretion disk-corona model, one needs very
short lived phenomena to occur above the disk's atmosphere. In fact, 
the whole evolution of the AR should happen faster than the disk's 
hydrostatic time scale. To confine the plasma
with a high compactness parameter $l\gg 1$, one needs mechanisms
other than gravitational confinement. We suggest that this points
to magnetic flares as the most likely mechanism for the AR formation.

\section{Magnetic Flares and Accretion Disks}

Galeev, Rosner \& Vaiana (1979) showed that magnetic
flares are likely to occur on the surface of an accretion disk,
since the internal dissipative processes are ineffective in
limiting the growth of magnetic field fluctuations. As a consequence
of buoyancy, magnetic flux should be expelled from the disk into a corona,
consisting of many magnetic loops, where the energy is stored.
It has also been speculated that just as in the Solar case, 
the magnetically confined, loop-like structures
(which we shall collectively call magnetic flares; see, e.g., Priest 1982) 
produce the bulk of the X-ray luminosity. 
The X-rays are assumed to be created by upscattering of the intrinsic
disk emission.

Since then, several Solar magnetic flare workers have elaborated on this
subject (e.g., Kuperus \& Ionson 1985; Burm 1986; Burm \& Kuperus 1988;
Stepinski 1991; de Vries \& Kuijpers 1992; Volwerk, van Oss \& Kuijpers 1993;
van Oss, van Oord \& Kuperus 1993). 
Unfortunately, these models are very much more complicated than simpler
plasma models that take into account the detailed interaction
of particles and radiation but leave out the question of how
the plasma is confined and energy is supplied. Thus, although the
models invoking magnetic flares above the cold accretion
disk have been viable, the detailed spectrum from such a flare 
could not be computed, and the model has remained somewhat of an abstraction.

An important step forward was that by Haardt, Maraschi \& Ghisellini
(1994), who for the first time attempted to connect
the physics of magnetic flares with the observational need for localized
active regions above the disk. However,  the 
actual consideration of the magnetic field structure that confines the 
plasma to the AR was still missing. Furthermore, 
the amount of energy stored in the magnetic field has been treated 
as just a parameter, depending on how long and at what rate the energy is supplied
to the AR. In reality, the field value is limited by the equipartition field 
in the disk (Galeev, Rosner \& Vaiana 1979). The question of how the pressure
equilibrium in the AR (important when discussing $\tau_T$ of the source) 
is set up has not been discussed. 

One of the purposes of this paper is to pay more attention to the magnetic
flare model for the X-ray emission from accretion disks 
in black hole systems in general, and in Seyfert Galaxies in particular.
In the rest of the paper, we point out that the MF model can account for many, 
if not all, of the observed X-ray and UV spectral features of Seyfert Galaxies.
Very importantly, we shall also demonstrate that these flares are
physically consistent with the constraints imposed on the ARs discussed 
above. 

\subsection{Possible Flare Geometry}

In the standard accretion disk theory, the gas density has an approximately
Gaussian vertical profile, and thus it decreases very fast with increasing height.
Let us also assume that the magnetic flux tube is rooted in the midplane 
of the disk. The ``flare region'', i.e., that part of the flux tube above 
the accretion disk surface, is then dominated by magnetic field pressure.
It is well known that a magnetic field, left to its own devices, tends
to fill all the available space (e.g., Parker 1979, \S 8.4).
For the magnetic flux tube rooted in the midplane of the disk, this means
that the tube cross section expands; the tube is thick in the sense that the
cross sectional radius is of the order of the tube length. The whole
structure has a roughly semi-spherical shape (Fig. 1). 

We note that the observations actually require the magnetic flux tubes to be
thick if they are to explain the X-ray emission from Seyferts. Indeed,
if the tubes are slim, then most of the photons reflected from the
disk will not re-enter the AR, but leave system. The amount of
cooling of the AR due to these photons is then not enough to explain the X-ray
indexes of Seyfert galaxies---from spectral modeling, it is known that
the fraction of photons re-entering the AR should be relatively
large, $\sim 1/2$ (e.g., Svensson 1996a). 

\vskip 0.2in
\begin{figure}[H] 
\centerline{\epsfig{file=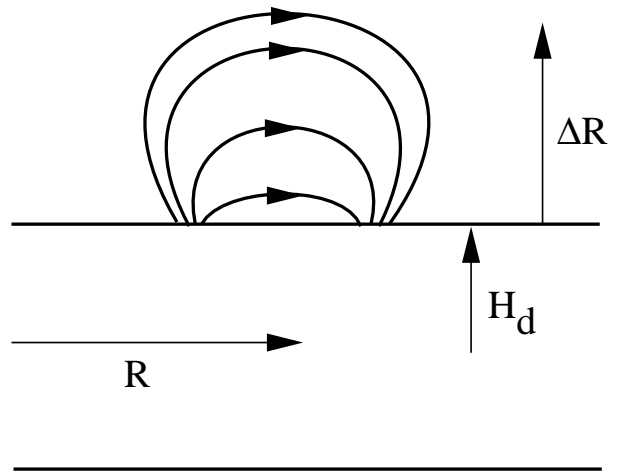,width=3.15in}}
\vspace{10pt}
\caption{Schematic of the magnetic flare geometry above the surface
of an accretion disk with scale height $H_d$.  The radius $R$ is
measured from the black hole, and the flux tube is here shown with a
vertical size ($\Delta R$) comparable to $H_d$.}
\label{fig1}
\end{figure}

\subsection{Maximum Compactness}

We will now assume that by some process (e.g., by magnetic reconnection
or dissipation of magnetic waves) the magnetic field
energy is being transferred to the particles. We can estimate the maximum
compactness of the AR by the following considerations. The magnetic field
is limited by the equipartition value in the midplane of the disk.
The size of the AR, $\Delta R$, is of the order of one turbulent cell, which
is at best of the order of the disk scale height $H_{\rm d}$.
Let us assume that the field annihilation (which provides the energy transfer to
the particles) occurs on a time scale $t_{\rm l}$ equal to the light
crossing time $H_{\rm d}/c$ times some number $b\gtrsim$ a few.
We will also assume that the flare occurs at 6 gravitational radii,
where most of the bolometric luminosity is produced. 
Using the results of SZ94, we obtain: 
\begin{equation}
l\leq \left({1\over b}\right) {\varepsilon_{\rm d} H_{\rm d}\over \Sigma_{\rm cr}} 
{\Delta R\over H_{\rm d}}.
\end{equation}
Here,  the `critical' column energy density is $\Sigma_{\rm cr}\equiv m_e c^2/\sigma_T$, where
$\sigma_T$ is the Thomson cross section and $\varepsilon_{\rm d}$
is the midplane energy density.  Using the results of SZ94, we obtain 
\begin{equation}
{\varepsilon_d H_d\over \Sigma_{\rm cr}}= C\times
(m_p/m_e)\,\alpha^{-1}r_3^{-3/2}{\cal L} J
\end{equation}
where $r_3\equiv 3R/R_g$ is the radius in units of 3 
gravitational radii, $R_g\equiv 2 G M/c^2$ 
($M$ is the mass of the black hole), $\alpha$ is the standard 
viscosity parameter, and ${\cal L}$ is the dimensionless
luminosity, ${\cal L} \equiv L/L_{\rm Edd}$, where
$L_{\rm Edd} = 2\pi (m_p c^2/\sigma_T R_g)$ is the Eddington luminosity.
Finally, $J\equiv 1-(1/r_3)^{1/2}$
accounts for the assumed stress-free boundary condition at 
the disk's inner edge.  The constant $C$ has the value $\sqrt{3/2}$ 
if the disk is gas pressure dominated, and it is $\sqrt{6}$ if 
the dominant pressure is due to radiation.

Taking $r_3 = 2$ as an example, we get 
\begin{equation}
l \lesssim 400 \, {{\cal L} \over \alpha b}\;
{\varepsilon_{\rm m} \Delta R
\over \varepsilon_{\rm d} H_{\rm d}}
\end{equation}
where $\varepsilon_{\rm m}$ is the magnetic energy density $\lesssim
\varepsilon_{\rm d}$. HMG94 suggested that plausible values
for $b$ and $\alpha$ are 10 and 0.1, respectively. We can also assume that
$\varepsilon_{\rm m}\sim 0.1 \varepsilon_{\rm d}$.
It is then seen that
$l\gg 1$, but it is not likely to be as high as a few hundred.

\subsection{Spectrum From Energetic Magnetic Flares}

The two-phase model is often criticized for a lack of self-consistency:
one of the most important quantities determining the spectrum---the
Thomson optical depth of the AR---is either fixed in an ad hoc manner, or
is said to be given by pair equilibrium. The latter may be viable if the
pairs are strongly confined inside the AR and if the 
compactness of the region is $\sim$ several hundred. However, a physical
description of how this happen is needed in order to validate the
basic assumptions of the model.  Haardt, Maraschi \& Ghisellini (1994) 
have made an attempt in this direction, but their description of 
magnetic flares was rather simplistic and did not provide an
explanation for the observed optical depth. To address this
issue in greater depth, Nayakshin \& Melia (1997b) considered 
the role played by pressure balance in establishing an equilibrium
optical depth during the active phase of a magnetic 
flare. The main difference with the Solar case is that here the 
compactness of the flare is much larger than unity,
and thus radiation pressure dominates over particle pressure (if the
proton temperature is the same as that of the electrons). The conditions
providing a pressure balance are therefore drastically different 
from those in the Sun, where the particles dictate the nature of the equilibrium.
Nayakshin \& Melia (1997b) assumed that the energy is supplied to the gas by
magnetohydrodynamic waves. Under the conditions typical for Seyfert
Galaxies, the group velocity of these waves ($v_a$) is  expected to be close
to the speed of light $c$. Because momentum is transferred to the gas,
as well as energy, a compressional force is imposed on the plasma. 
The radiation pressure within the active region is approximately
$\tau_T F_x/c$, where $F_x$ is the X-ray flux leaving the source.
In quasi-equilibrium the energy influx is equal to the energy outflux, and
radiation pressure is equal to the momentum influx due to the magnetic waves.
This then requires that the Thomson optical depth $\tau_T$ be in the range
$0.5 - 2$, depending on the actual geometry of the flare. 

The Alfv\'en velocity can be used as an estimate for the group velocity 
of the magnetohydrodynamic waves.  Taking the disk structure to be that of
a standard Shakura-Sunyaev model in its radiation pressure dominated region,
one can show that the Alfv\'en velocity $v_a$ (at a distance of $6$ 
gravitational radii from the black hole) is
\begin{equation}
{v_a\over c}\simeq {1\over 2}\,\left [{\varepsilon_m\over \varepsilon_d}
\alpha^{-1}{\cal L}\tau_T^{-1}\right]^{1/2}\;.
\end{equation}
It is evident that for $\varepsilon_m/\varepsilon_d$ 
and ${\cal L}$ not too small, $v_a$ can be quite
close to $c$ (if it exceeds $c$, the relativistic corrections will
permit it to saturate at $c$ only).  In this estimate, we assume that the 
Thomson optical depth $\tau_T$ of the plasma within the flare region
is entirely due to the accreting electrons. If in addition
pairs are produced, then Equation (6) should be used with $\tau_p$ instead, 
where $\tau_p$ is the Thomson optical depth of the AR due to the 
electrons accreting with the protons, which further increases $v_a/c$.
We conclude from this that $v_a$ must be close to $c$ for quite a broad range of
the $\alpha$-parameter, $\cal{L}$, and it is {\it completely independent} of 
the black hole mass $M$.

As already noted by Haardt, Maraschi \& Ghisellini (1994), the spectrum
of a magnetic flare should be similar to that of a static
active region of the same size and compactness, as long as the lifetime of
the flare exceeds several light-crossing time scales. This is certainly true
if pairs are not important, since the time scales for other processes 
that may influence the spectrum (e.g., Poutanen \& Svensson 1996a) are of the
order of a light crossing time. However, the life time of one single
flare is short compared with the typical integration time of current X-ray
instruments. Moreover, it is very likely that there are many magnetic
flares present at any given moment of time. Therefore, it becomes clear
that if magnetic flares are responsible for the X-ray emission from 
Seyfert Galaxies, the spectrum must be a composite of the contributions
from many different flares.  Nayakshin \& Melia (1997b) tested this 
possibility, assuming that the energy balance is fixed by requiring 
the Compton $y$-parameter to be constant for all the flares (which is 
reasonable, given that $y$ is fixed by the geometry of the two-phase 
model), and they summed over the spectra from flares with different 
$\tau_T$. For illustrative purposes, the
distribution of flares was taken to be a Gaussian over $\tau_T$,
centered on $1.14$ with a dispersion of $0.7$. The resulting
spectrum is practically indistinguishable from that of a single flare 
with $\tau_T = 1.14$ up to a photon energy of about a hundred keV.
The OSSE error bars are much larger than the deviations of the composite 
and single flare spectra, and so the current observations cannot
distinguish between these two possibilities.  Thus, magnetic flares 
can conceivably account for the observed X-ray/$\gamma$-ray 
spectra of Seyfert Galaxies.

\subsection{Explanation of the BBB Temperature}
 
Nayakshin \& Melia (1997c) considered the X-ray reflection/reprocessing
due to a transient, energetic flare above the accretion
disk to compare with other studies reported in the literature that assume
a stationary state. The main difference between the two is
the structure of the emitting (i.e., reprocessing) layer. 
In particular, since the flare lifetime is shorter than the disk thermal
time scale, a pressure and energy equilibrium between the 
incident X-ray flux and the underlying disk is not established.
A typical photon does not have sufficient time to diffuse to
the mid plane of the disk during one lifetime of the flare.
However, the X-ray skin, i.e., the layer that absorbs and reprocesses
the X-rays, is only a tiny fraction of the whole disk, and thus a
quasi-equilibrium is established within it.  As a result of the incident flux,
the X-ray skin is compressed to much higher densities
than the density of the undisturbed accretion disk. It turns out that
the pressure and energy equilibrium of this X-ray skin yields 
a unique temperature $\sim $ few $\times 10^5\,$K {\it independently}
of the mass of the central engine. This seems to account well
for the observed independence of the Big Blue Bump temperature on the luminosity of
the source (Walter 1994; Zhou et al. 1997).
By comparison, a stationary, time independent reflection cannot easily
explain these observations.

An additional attractive feature of the MF model
is that due to a much larger gas density in the reflecting layer,
the ionization parameter ($\xi \sim 20$) remains relatively small,
in which case the reflected/reprocessed spectrum is 
indistinguishable from that of a neutral reflector, which appears to
be favored by current observations (Zdziarski et al. 1996). 
Static X-ray reflection/reprocessing, on the other hand, may have 
difficulties complying with the observed low ionization parameter 
of the reflecting matter, since in this case
the X-ray skin density is much lower. Summarizing, many of
the attractive features of reflection/reprocessing in a static
layer below the AR are preserved in the case of a time-dependent,
short-lived magnetic flare, but the latter has the additional
advantage of being able to account for the approximate universality of
the BBB temperature and the low ionization fraction in the reflector.

\subsection{Pair Equilibrium within the Magnetic Flares}

One of the central questions in the modeling of Seyfert Galaxies
has always been whether a pair equilibrium is established within the source,
since this has some serious observational consequences.  However,
pairs have successfully eluded detection in Seyfert Galaxies.
With the discovery of a high-energy break above $\sim$ 100 keV
and the non-detection of a predicted annihilation line, it has
become apparent that the non-thermal power in Seyfert Galaxies, 
if at all present, is quite small (e.g., Svensson 1996a; 
Zdziarski et al. 1996, and references therein).
Thus, it was concluded that the plasma is mostly thermal (e.g.,
Haardt \& Maraschi 1991; Fabian 1994).
This inference was supported by the finding that
an annihilation line would not be observed from a thermal 
plasma because it is always hidden in the broad Comptonized spectrum
(Zdziarski \& Coppi 1995).

Recent work by (Zdziarski et al. 1996) suggests that in the
context of a thermal pair equilibrium, an optical depth
of roughly unity is then the consequence of a large
compactness ($\sim$ several hundred). We, however, suggest that this
situation is achieved by pressure equilibrium, as discussed in
\S 3.3. In this case, the plasma consists primarily of the electrons
and protons stripped from the disk, at least at 
the beginning of the flare, since during the
magnetic energy storage phase the plasma is not sufficiently hot to provide
enough hard photons that would create electron-positron pairs.
Thus, in this framework, the pairs are not important in determining the
spectrum from the flare, and this is again consistent with the
lack of any observed pair signature.

Of course, a detailed modeling of a magnetic flare event must take into
account the pair creation process which continuously 
produces new pairs when $l\gtrsim 10$.  It is the total optical depth
(i.e., the sum of the Thomson optical depths of electrons and pairs) that
matters for the pressure equilibrium.  If this pressure balance fixes the 
optical depth to some particular value $\sim 1$, then clearly, compared to the
no-pair case, the plasma must expand to accommodate the new particles. 
Let us assume that the total energy supplied to the plasma is a constant,
which means that the luminosity $L$ remains constant. Then, as the plasma
expands, its compactness decreases as $1/\Delta R$ since $l\sim L/\Delta R$. 
Since the pair creation rate is proportional to $l^2$, an
equilibrium is reached at some $\Delta R$ such that the pairs are now responsible for
a fraction of the total optical depth $\tau_T$. This fraction
turns out to be quite small unless the initial compactness is
as high as several hundred. It is interesting to note that even flares with 
an initial value of $l$ that would lead to a
pair runaway (e.g., Svensson 1982) find an equilibrium configuration with
a source compactness below this critical value.
We intend to quantify the character of the pair equilibrium 
in this situation in a future publication, but we may
already anticipate that a compactness as high as several hundred is only 
barely permitted by Equation (5), and that 
therefore pairs should be of relatively low importance to 
the dynamics and energetics of magnetic flares.

\subsection{Magnetic Flares and AGN Light Curves}

Several authors have suggested that magnetic
flares above the accretion disk are responsible for the observed variations
in the AGN and GBHC luminosity (e.g., Galeev, Rosner \& Vaiana 1979;
de Vries \& Kuijpers 1992; Volwerk, van Oss \& Kuijpers 1993;
van Oss, van den Oord \& Kuperus 1993, and others).   The power
density spectrum (PDS) from these sources is typically a power-law
(Lawrence et al. 1987; McHardy \& Czerny 1987; Krolik et al. 1991).
In the case of the Sun, Dmitruk \& Gomez (1997) have shown that 
magnetic flares can naturally account for a power-law shape in
the PDS with an index $\simeq 1.5$.  Since in principle the 
flares in black hole systems may have different spatial sizes,
and thus different durations and overall power, one can reasonably
expect that a similar PDS may be produced by these transient events
above the accretion disks in AGNs and GBHCs. 

We note here that the power-law PDS should be explained by 
local variations of the magnetic flare properties, rather than
variations occurring systematically with a changing location of the flare
(compare with the rotating bright-spots model, e.g., Abramovicz
et al. 1991). The observed X-ray PDS spans a wide range in frequencies,
typically $10^{-5}$-$10^{-3}$ Hz. This range corresponds to the range
in radius $\sim 30$, since $\Omega^{-1}\sim R^{3/2}$, where 
$\Omega$ is the rotational frequency of the Keplerian disk. 
But the local contribution to the overall luminosity goes as 
$J/R^2$, and thus the smallest frequencies contribute less than
the largest ones, in contradiction to the observed power spectrum.
Only if one assumes that the luminosity of the flare is independent of its location
does one obtain the right power spectrum. However, such an assumption is
unphysical, since we know that the X-ray luminosity is a major
component of the bolometric luminosity, and thus it should scale 
in the same way as the local gravitational dissipation in the disk.

Therefore, since the emission comes from a relatively 
narrow range in radii, it should be the flare size that varies 
and produces the observed PDS.
Alternatively, since disturbances propagate along magnetic field
lines in a strong magnetic field, and since the magnetic flux tube
is thick, there can be a wide range in characteristic 
scales $D$ even in one source ($D$ is essentially the length of the given
magnetic field line [see Fig. 1]). Moreover, the energy density
of the magnetic field will scale roughly as $1/D^2$ (that would
be so for a potential field that has no currents even at the boundary,
i.e. in the footpoints). Thus, one might expect to see a power-law 
PDS even from a {\it single} event in this case. We intend to investigate this
question in future work,  but we caution that the analysis of the 
PDS is unlikely to provide any valuable information about one single
magnetic flare, since at any given instant of time there should be
a number of such events. These flares occur roughly at random, 
and thus information about a single flare is washed out.

The complete annihilation of the 
magnetic field energy $\varepsilon_{\rm m}$ ($\approx
\varepsilon_{\rm d}$) within a volume $H_d^3$ during
a time $b\, H_d/c$ provides an estimate of the single flare luminosity:
\begin{equation}
{L_1 \over L}\,\lesssim { {\cal L}\over 4 b}\;,
\end{equation}
where we have used the SZ94 accretion disk parameters
with their $f$ set equal to $1/2$. Based on similar considerations, 
HMG94 estimated the required number of magnetic flares to
be about 10. We are therefore in agreement with this estimate, 
although in principle the number of less energetic or smaller
flares may still be larger , since Equation (7) is only an upper limit
on $L_1$.

\subsection{Gravity Constraints}

An implicit assumption thus far has been that the magnetic flare can 
indeed sustain a sufficient number of protons roughly one disk height 
$H_d$ above the disk. For this to be viable, we need to demonstrate
that the gravitational energy of the protons trapped inside the flux tube is
very much smaller than the magnetic field energy.
The latter is at most the total thermal energy of the disk immediately below
the flare, while the former may be estimated as $E_{\rm grav}\sim 
3^{-1}\,(\tau_T/3 r_3)(m_pc^2/\sigma_T)(H_d/R)^2 H_d^2$. Using 
expressions from SZ94 for $r_3 =2$, we see that 
\begin{equation}
{E_{\rm grav}\over \varepsilon_d H_d^3} \simeq
2\times 10^{-2} c^{-1} \alpha \tau_T {\cal L}\ll 1\;.
\end{equation}
which satisfies the constraint.  As before, $\tau_T\simeq 1$ 
is the Thomson optical depth of the material trapped inside the flux tube.

\subsection{Stability of the Accretion Disk}

The nature of accretion disk instabilities has received a great
deal of attention (for recent references, see Chen 1995).  While 
not attempting to consider this question in detail here, we can
make several comments on the stability of the MF model.

Magnetic flares may be viewed as an additional channel by which
energy can be transported out of the disk.  Of course, in the standard
disk model, the dissipated gravitational energy is lost directly to radiation. 
Since the time taken by a photon to diffuse outward from the midplane 
to the disk's surface is a very strongly increasing  
function of the optical depth, it is conceivable that under some
conditions the energy transported by the magnetic field is 
greater than that due to the radiation.  The total energy content
of the disk plus corona system is then expected to be lower than
that of the standard theory, though with the same luminosity,
and such a situation leads to greater stability (e.g., SZ94).
Although it is not clear what role advection would have in such
a model, it is expected that magnetic flares may help to quench
some of the disk instabilities encountered in standard models.

\subsection{Remaining Questions and Problems}

We have seen that magnetic flares are physically consistent with the
multi-wavelength spectra of Seyfert Galaxies.  Very importantly,
the MF model seems to account for several observed characteristics
that cannot be easily reconciled with a picture in which the ARs are
static. However, a host of unanswered questions and problems remain.

First, accretion disk flares have been considered only in a highly
schematic fashion thus far.  Unfortunately, the physics of 
magnetic energy release in a non-static and turbulent gas is not well known, other
than the fact that it must happen, as is seen in the Sun. In addition,
a detailed model of the magnetic flare should also include a consideration of
all the relevant aspects of magnetic flux tube formation in the underlying
turbulent disk, a problem that also has not been solved. This, however,
does not mean we can ignore the magnetic flare model for the X-ray emission
in Seyferts.  Instead, additional studies are called for,
especially in view of the fact that very recent observations of Solar flares
seem to support much of the current theoretical thinking in this
area and are generating enthusiasm among solar theorists (e.g., 
Innes et al. 1997, Klimchuk 1997).

Another major unresolved issue is how the disk viscosity is connected
to the magnetic field. If we knew this relationship, we would be able to
eliminate $\alpha$ or $\varepsilon_{\rm m}$ from Equation (5), and thus get 
better constraints on the maximum compactness of a magnetic flare.  This
follows from the fact that the structure of a cold disk is quite sensitive 
to the viscosity law.  In addition, viscosity figures very prominently 
in the physics of magnetic flux tubes (e.g., Vishniac 1995). 

\section{Conclusions}

In this paper we have attempted to address the problems that arise
when physical constraints are imposed on the active regions thought to exist
in the two-phase corona-accretion disks in Seyfert Galaxies.
We showed that these regions should necessarily be highly transient,
i.e., evolve faster than one thermal disk time scale due to spectrum
formation constraints. A consideration of the plasma confinement lead us to
require an overall magnetic field with a stress much larger than 
the X-ray radiation pressure. Furthermore, putting these constraints together,
we concluded that the magnetic flare model appears to be consistent with
the type of transient active regions required by the observations.
We then proceeded to show that the model is probably capable of explaining the observed
optical depth, X-ray reflection and UV reprocessing implied by the data,
and the observed power-law power density spectra. Finally, we discussed
the unresolved issues that need to be investigated in future work.

\section{Acknowledgments}

This work was partially supported by NASA grant NAG 5-3075.  We have
benefitted from many discussions with Randy Jokipii and Eugene Levy.

{}

\end{document}